\documentstyle[multicol,aps,epsf]{revtex}

\draft

\begin{document}

\title{\bf Critical Behavior of Damping Rate for Plasmon\\
        with Finite Momentum in $\phi^4$ Theory}

\author{Hanzhong Zhang, Luan Cheng and Enke Wang}

\address{Institute of Particle Physics, Huazhong Normal University,
 Wuhan430079, China}

\date{April 4, 2004}

\maketitle

\begin{abstract}

Applying thermal renormalization group (TRG) equations to $\phi^4$
theory with spontaneous breaking symmetry, we investigate the
critical behavior of the damping rate for the plasmons with finite
momentum at the symmetry-restoring phase transition. From the TRG
equation the IR cutoff provided by the external momentum leads to
that the momentum-dependent coupling constant stops running in the
critical region. As the result, the critical slowing down
phenomenon reflecting the inherently IR effect doesn't take place
at the critical point for the plasmon with finite external
momentum.
\end {abstract}


\begin{multicols}{2}

As well known, at high temperature the spontaneous-breaking
symmetry in scalar field theory can be restored through phase
transition. At finite temperature the effective degrees of the
freedom of the scalar field are collective modes which is
interpreted as quasiparticles, the so-called plasmons. The plasmon
possess finite thermal mass which is generated dynamically by the
interactions among the fundamental degrees of freedom of field.
The thermal mass plays an important role at the phase transition
for restoring spontaneous-breaking symmetry. The width of the
spectral density of the plasmon is described by damping rate
defined as\cite{Weldon}
\begin{eqnarray}
   \gamma_{\bf{k}}(T)\equiv\frac{{\rm Im}\Sigma(\omega_{\bf{k}},\bf{k})}
                       {2 \omega_{\bf {k}}}\, ,
\end{eqnarray}
where ${\rm Im}\Sigma$ is the imaginary part of the self energy,
$\omega_{\bf k}^2=|{\bf k}|^2+m^2_p(T)$, $m_p(T)$ the plasmon mass
and ${\bf k}$ the plasmon momentum. As shown by Weldon in
Ref.\cite{Weldon}, if the plasma is slightly out of thermal
equilibrium, then $\gamma_{\bf k}(T)$ gives half the relaxation
rate (or the inversion of the relaxation time) of the
quasiparticle distribution function to its equilibrium value,
\begin{equation}
   \frac{d\delta n_{\bf k}}{dt}=-2\gamma_{\bf k}(T)
                   \delta n_{\bf k}\nonumber\, ,
\end{equation}
where $\delta n_{\bf k}$ is the deviation of the distribution
function from equilibrium, $\delta n_{\bf k}=n_{\bf k}-n_{\bf
k}^{eq}$.

In Refs.\cite{Parwani} and \cite{Jeon} Parwani and Jeon
investigated the damping rate of the plasmon at rest in massless
$\phi^4$ theory. In our previous paper\cite{Wang}, we generalized
their work to discuss the damping rate of plasmon with finite
momentum. All these works don't take into account the effect of
thermal renormalization group on the coupling constant. As shown
by Pietroni\cite{Pietroni}, for plasmon at rest the damping rate
is divergent at critical temperature of the phase transition for
restoring symmetry in $\phi^4$ theory with spontaneous-breaking
symmetry. This critical behavior contradicts the critical slowing
down law. The singularity of the damping rate of the plasmon at
critical point can be cured by taking into account the running
coupling constant with temperature from thermal renormalization
group (TRG) equation. In this paper, we will generalize Pietroni's
work to investigate the critical behavior of the damping rate for
the plasmons with finite momentum in general case.

We consider following Lagrangian in $\phi^4$ theory  with
spontaneous-breaking symmetry,
\begin{equation}
   {\cal L}_0 =
   \frac{1}{2}{\partial^\mu\phi}{\partial_\mu\phi}
   +\frac{1}{2} \mu_0^2\phi^2+\frac{\lambda_0}{4!}\phi^4\, ,
\label{Lag}
\end{equation}
where $\mu_0^2<0$ and $\lambda_0$ is coupling constant. A
consistent determination of the plasmon damping rate requires the
resummation of hard thermal loop\cite{Pisarski}. The induced
thermal mass correction resulting from the hard thermal loop (the
tadpole diagram) reads
\begin{equation}
    \Delta m^2_T=\frac{\lambda_0T^2}{24}\nonumber\, .
\end{equation}
The plasmon mass can be defined as
\begin{eqnarray}
   m_p^2=\mu_0^2+\frac{\lambda_0T^2}{24}\, .
\end{eqnarray}
Notice that the thermal mass correction $\Delta m^2_T$ is
positive. Although $\mu_0^2<0$, at enough high temperature $T>T_c$
the plasmon mass becomes positive, the spontaneous-breaking
symmetry is restored. The critical temperature $T_c$ can be
expressed as
\begin{equation}
  T_c=\sqrt{-\frac{24\mu_0^2}{\lambda_0}} \nonumber\, .
\end{equation}

In $\phi^4$ theory, the resummation of the hard thermal loop is
much easier since the hard thermal loop is just a
momentum-independent real constant. The effects from hard thermal
loop can be resummed by defining an effective Lagrangian
through\cite{Wang}
\begin{eqnarray}
    {\cal L} &=&({\cal L}_0+\frac12 \Delta m^2_T\phi^2)
             -\frac12 \Delta m^2_T\phi^2
   \nonumber\\
           &=&{\cal L}_{eff}-\frac12 \Delta m^2_T\phi^2\, ,
\end{eqnarray}
and treating the last term as an additional interaction. This
effective Lagrangian defines an effective propagator
\begin{equation}
   {\Delta(k)}=\frac1{k^2+\mu_0^2+\Delta m^2_T}\nonumber\, .
\end{equation}

The leading contribution to the imaginary part of the self-energy
comes from the two loop sunset diagram. It can be expressed as
\cite{Wang}
\begin{eqnarray}
  {\rm Im}\Sigma_2(\omega_{\bf p},{\bf p})={\rm Im}\Sigma^{3BD}(\omega_{\bf p},{\bf p})
        +{\rm Im}\Sigma^{LD}(\omega_{\bf p},{\bf p})\, .
\end{eqnarray}
${\rm Im}\Sigma^{3BD}$ and ${\rm Im}\Sigma^{LD}$ are contribution
from the 3-body decay and Landau damping, respectively:
\begin{eqnarray}
   {\rm Im}\Sigma^{3BD}(\omega_{\bf p},{\bf p})&=&\pi (e^{\frac{\omega_{\bf p}}{T}}-1)
        \int d[{\bf k}, {\bf q}]f_{\bf k}f_{\bf q}f_{\bf r}
          \nonumber\\
          &~&\times
    \delta (\omega_{\bf p}
        -E_{\bf k}-E_{\bf q}-E_{\bf r})\, ;
\\
     {\rm Im}\Sigma^{LD}(\omega_{\bf p},{\bf p})&=&3\pi (e^{\frac{\omega_{\bf p}}{T}}-1)
     \int d[{\bf k}, {\bf q}](1+f_{\bf k})f_{\bf q}f_{\bf r}
         \nonumber\\
          &~&\times
         \delta (\omega_{\bf p}
      +E_{\bf k}-E_{\bf q}-E_{\bf r})\, ,
\end{eqnarray}
with the same notation as in Ref.\cite{Wang},
\begin{eqnarray}
   d[{\bf k}, {\bf q}]&=&\frac{\lambda^2\mu^{4\epsilon}}{6}
          \frac{d^{D-1}{\bf k}}{(2\pi)^{D-1}}
  \frac{d^{D-1}{\bf q}}{(2\pi)^{D-1}}\frac{1}{8E_{\bf k} E_{\bf q} E_{\bf r}}
\\
   {\bf r}&=&{\bf{k}}+{\bf{q}}-{\bf{p}}\, ,
\\
   E_{\bf l}^2&=&{\bf l}^2+m_p^2\, ,\qquad {\bf l}={\bf k},{\bf q},{\bf r}\, ,
\\
   f_{\bf l} &=&\frac{1}{\exp{(E_{\bf l}/T)}-1}\, , \qquad {\bf l}={\bf k},{\bf q},{\bf r}\, .
\end{eqnarray}

The plasmon damping rate with finite momentum can  be obtained
as\cite{Wang}
\begin{eqnarray}
   \gamma (\sqrt{|{\bf p}|^2+m_p^2}, {\bf p})
    =\frac{\lambda_0^2T}{256\pi^3}\frac{1}{z\epsilon(z)}f(z)\, ,
\label{rate}
\end{eqnarray}
where
\begin{eqnarray}
   f(z)&=&\int^z_0 dx\left [L_2(\xi)+L_2\left (
     \frac{\xi-\zeta}{\xi(1-\zeta)}\right)\right.
\nonumber\\
    &-&\left. L_2 \left(
    \frac{\xi-\zeta}{1-\zeta}\right)
   {-}L_2\left( \frac{(\xi-\zeta)(1-\xi\zeta)}{\xi(1-\zeta)^2}\right)
    \right],
\\
     z&=&\frac{|{\bf p}|}{T},\quad a=\frac{m_p}{T},\quad
       \epsilon(z)=\sqrt{z^2+a^2},
\\
   \xi&=&e^{-\epsilon(z)},\quad\zeta=e^{-\epsilon(x)}\, ,
\\
     L_2(y)&=&-\int^y_0 dt\frac{\ln (1-t)}{t}\, .
\end{eqnarray}

For the plasmons at rest (${\bf p}=0$) the damping rate reduces to
\begin{eqnarray}
    \gamma (m_p^2, {\bf p}=0)=\frac{\lambda_0^2T^2}{1536\pi m_p^2}\, .
\label{rate0}
\end{eqnarray}
As $T\rightarrow T_c$ the vanishing plasmon mass results in
divergent damping rate for the plasmons at rest. This means that
the relaxation time becomes shorter and shorter as the critical
temperature is approached. This behavior contradicts the critical
slowing down law exhibited in the condensed matter systems. We
should notice that, in getting above result, the coupling constant
$\lambda_0$ is considered as a temperature-independent. Actually,
in thermal field theory the coupling constant runs with
temperature from thermal TRG equation.

As shown in Ref.\cite{Attanasio}, in the framework of Wilson
renormalization group, the TRG equation in real time thermal field
theories is deduced as
\begin{equation}
   \Lambda \frac{\partial\Gamma_\Lambda[\varphi]}{\partial \Lambda}
   {=}
   \frac{i}{2}
   {\mbox{Tr}} \left[ \Lambda \frac{\partial D_\Lambda^{-1}}{\partial \Lambda}
    {\cdot}\left( D_\Lambda^{-1} {+} \frac{\delta^2 \Gamma_\Lambda
    [\varphi]}{\delta\varphi\delta \varphi} \right)^{-1} \right],
\label{TRG}
\end{equation}
where $\Lambda$ is a cut-off introduced in the thermal sector of
real-time propagator in Closed Time Path (CPT)
formalism\cite{Schwinger} by modifying the Bose-Einstein
distribution function $N(k_0)=1/[\exp(k_0/T)-1]$ as
\begin{equation}
  N_{\Lambda}(k_0)=N(k_0)\theta(|{\bf k}|-\Lambda)\, ;
\label{Nlam}
\end{equation}
$D_{\Lambda}$ is the tree level propagator with $N_{\Lambda}(k_0)$
in the CPT formalism; $\Gamma_{\Lambda}$ is the generating
function of 1PI vertex function in which the modes with
$k>\Lambda$ have been integrated out.

From Eq.(\ref{TRG}) the evolution equations for 2-point and
4-point functions can be expressed as
\begin{eqnarray}
   \Lambda\frac{\partial}{\partial\Lambda}\Gamma_\Lambda^{(2)}
   &=&\frac{1}{2} {\rm Tr}[K_\Lambda
   \Gamma_\Lambda^{(4)}]\, ,
\label{Lam2}\\
   \Lambda\frac{\partial}{\partial\Lambda}\Gamma_\Lambda^{(4)}&=&
   -3{\rm Tr}[G_\Lambda \Gamma_\Lambda^{(4)} K_\Lambda \Gamma_\Lambda^{(4)}]\, ,
\label{Lam4}
\end{eqnarray}
where $K_{\Lambda}$ and n-point 1PI vertex function are defined as
\begin{eqnarray}
   &&K_\Lambda(k,\varphi)\equiv
   -iG_\Lambda\cdot\Lambda\frac{\partial}{\partial\Lambda}
     D_\Lambda^{-1}\cdot G_\Lambda\, ,
\label{kG}\\
    &&\Gamma_{\Lambda}^{(n)}(k,\varphi)=\frac{\delta^n\Gamma_\Lambda[\phi]}
    {\delta\phi_{i_1}\delta\phi_{i_2}\cdot\cdot\cdot\delta\phi_{i_n}}
    |_{\phi_{i_1}=\cdots=\phi_{i_n}=\varphi}\, .
\label{kn}
\end{eqnarray}
Here $G_\Lambda$ is the full propagator
\begin{equation}
     G_\Lambda^{-1}(k; \varphi)=D_\Lambda^{-1}(k; \varphi)
                +\Sigma_\Lambda(k; \varphi)\, ,
\label{G1}
\end{equation}
self-energy $\Sigma_\Lambda$ is 1PI 2-point vertex,
\begin{equation}
   \Sigma_\Lambda(k; \varphi)=\Gamma_\Lambda^{(2)}(k; \varphi)
   \equiv\frac{\delta^2\Gamma_\Lambda [\phi]}
   {\delta \phi \delta \phi}|_{\phi_1 =\phi_2 =\varphi}\, .
\label{enself}
\end{equation}

Define 4-point function  and cut-off mass as
\begin{eqnarray}
    \Gamma_\Lambda^{(4)}&=&-\lambda_\Lambda -i\eta\, ,
\\
    m^2_\Lambda &=& \mu_0^2-{\rm Re}\Gamma^{(2)}_\Lambda\, ,
\end{eqnarray}
and substitute them into Eqs.(\ref{Lam2}) and (\ref{Lam4}), the
TRG equations for the plasmon mass and the coupling constant are
deduced as
\begin{eqnarray}
  \Lambda\frac{\partial}{\partial\Lambda}m_\Lambda^2
    &=&-\frac{\Lambda^3}{4\pi^2}\frac{N(\omega_\Lambda)}
        {\omega_\Lambda}\lambda_\Lambda\, ,
\label{m1}\\
  \Lambda\frac{\partial}{\partial\Lambda}\lambda_\Lambda
   &=&-3\frac{\Lambda^3}{4\pi^2}
    \left(\frac{d}{dm_\Lambda^2}\frac{N(\omega_\Lambda)}
       {\omega_\Lambda}\right)\lambda_\Lambda^2\, ,
\label{m2}
\end{eqnarray}
where $\omega_\Lambda=\sqrt{\Lambda^2+m^2_{\Lambda}}$.

The initial conditions for the evolution equations (\ref{m1}) and
(\ref{m2})are given at a scale $\Lambda =\Lambda_0\gg T$,
\begin{equation}
   m^2_{\Lambda_0}=\mu^2_0\, , \qquad
   \lambda_{\Lambda_0}=\lambda_0\, .
\end{equation}
Due to the exponential suppression in the Bose-Einstein
distribution function, it is enough to take $\Lambda_0=10T$.

 \begin{figure}
 \begin{center}
 \epsfxsize 66mm \epsfbox{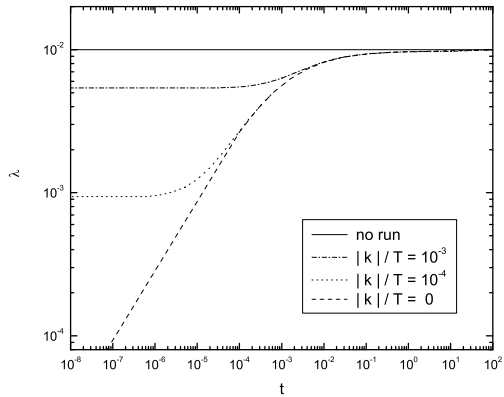}
 \end{center}
 \caption{\label{fig1}
 \small The running of the coupling constant versus dimensional variable $t=(T-T_c)/T_c$
        given by the TRG equations. The dashed, dotted and dash-dotted curves are for
        $|{\bf k}|/T$=0, $10^{-4}$ and $10^{-3}$, respectively. The solid line is for
        no-running coupling constant fixed to 0.01.}
\end{figure}

For the moving plasmon with finite momentum $|{\bf k}|$, the
non-vanishing external momentum $|{\bf k}|$ provides an IR cutoff
to the TRG running, so the running will effectively stop as
$\Lambda$ is of the order of $|{\bf k}|$. When $\Lambda\rightarrow
|{\bf k}|$, we take the solutions $m^2_{\Lambda =|{\bf k}|}$ and
$\lambda_{\Lambda =|{\bf k}|}$ as the results we are expecting
for. For the plasmon with zero and finite external momenta, the
coupling constants $\lambda_{\Lambda =|{\bf k}|}$ versus
dimensionless variable $t=(T-T_c)/T_c$ are illustrated in
Fig.\ref{fig1}. The numerical results show indeed that, because of
the IR cutoff provided by the external finite momentum, the
momentum-dependent coupling constant stops running with
temperature and keeps a constant in the critical region. As shown
in Fig.\ref{fig1}, the region keeping the momentum-dependent
coupling constant as a no-running constant increases with
increasing the external momentum. In the following, we will show
that the above features of the running coupling constant will
change the critical behavior for the plasmon with finite external
momentum at the critical point by comparing to that for the
plasmon with vanishing external momentum.

\begin{figure}
 \begin{center}
 \epsfxsize 66mm \epsfbox{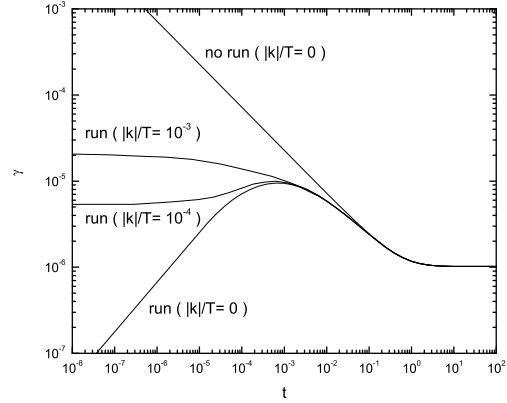}
 \end{center}
 \caption{\label{fig2}
 \small The plasmon damping rate versus dimensional variable $t=(T-T_c)/T_c$ for different
        momenta. The upper
        and lower curves are for plasmon at rest with no-running and running coupling
        constant, respectively. The middle two curves result from the momemtum-dependent
        running coupling
        constant for the plasmon with
        momentum $|{\bf k}|/T=10^{-3}, 10^{-4}$, respectively.}
\end{figure}

Substituting running coupling constant into Eq.(\ref{rate}), we
get the damping rate versus temperature as illustrated in
Fig.\ref{fig2}. For the damping rate of the plasmon at rest, the
result in Ref.\cite{Pietroni} is reproduced by the lower curve in
Fig.\ref{fig2}. It is clear that the damping rate goes to zero
instead of infinity (divergence) as temperature approaches to
critical point, by taking into account the running coupling
constant with temperature from TRG equation. This corresponds to
that the relaxation time goes to infinity at critical point, the
behavior is consistent with the critical slowing down law. For the
plasmons with finite momentum the damping rates versus temperature
are illustrated by the middle two curves in Fig. \ref{fig2} for
the external momentum $|{\bf k}|/T=10^{-3}$ and $10^{-4}$,
respectively. For much small external momentum $|{\bf
k}|/T=10^{-4}$, the momentum-dependent coupling constant can run
into the critical region, the damping rate goes up at first as
temperature approaches gradually the critical region, and then
goes down with coupling constant running down as the temperature
enters the critical region. After entering the critical region,
tendency of the plasmon damping rate is opposite to that obtained
from no-running coupling constant, and the relaxation time gets
longer and longer, which is then consistent with the critical
slowing down law. As the critical point is approached further, the
momentum-dependent coupling constant stops running with
temperature, the plasmon damping rate stops decreasing and doesn't
vanish, the critical slowing down phenomenon doesn't occur at the
critical point. As $|{\bf k}|/T=10^{-3} $, the momentum-dependent
coupling constant can't run into the critical region. Although the
damping rate $\gamma_{\bf k}(T)$ is no longer divergent as the
critical temperature is approached, which is different from the
critical behavior of the plasmon with zero external momentum for
no-running coupling constant, but the damping rate increases still
with approaching the critical temperature. This means that the
relaxation time gets shorter and shorter as temperature approaches
critical point. the critical slowing down phenomenon doesn't take
place completely. We notice that the critical slowing down
phenomenon is an inherently IR effect, which takes places only in
the double limit for external momentum $|{\bf k}|\rightarrow 0$
and the cut-off $\Lambda\rightarrow 0$.

In summary, based on the TRG equations in the CTP formalism, we
investigate the critical behavior of the damping rate of the
plasmon with finite momentum in $\phi^4$ theory with spontaneous
breaking symmetry. For the plasmon with vanishing external
momentum, the no-running coupling constant leads to that the
critical behavior is against the critical slowing down law, this
violation can be cured by taking into account the running of the
coupling constant with temperature from the TRG equations.  For
the plasmon with finite external momentum, the IR cutoff provided
by the external finite momentum leads to that the
momentum-dependent coupling constant stops running with
temperature in the vicinity of the critical point. Only for enough
smaller external momentum, the momentum-dependent coupling
constant can run into the critical region, we can see the tendency
of critical slowing down phenomenon. Nearby the critical point,
the momentum-dependent coupling constant stops running and keeps a
constant, the critical slowing down phenomenon doesn't take place
for the plasmon with finite external momentum because the critical
slowing down phenomenon is an inherently IR effect. Our recent
papers show that the shear viscosity of the thermal scalar field
is closely related to the damping rate of the plasmons with finite
momentum\cite{Wang2}. We anticipate that, by taking into account
the TGR equation, the damping rate discussed in this paper will
result in important effects on the transport properties of the
thermal scalar field. The further work has been in progress.

This work was supported by the National Natural Science Foundation
of China under Grant Nos. 10440420018 and 10135030. H.Z.Z.thanks
M. Pietroni for helpful discussions by e-mails.

\end{multicols}
\end{document}